# High-throughput Biological Cell Classification Featuring Real-time Optical Data Compression

Bahram Jalali[*], Ata Mahjoubfar, and Claire L. Chen

Department of Electrical Engineering, University of California, Los Angeles, California 90095, USA
**Plenary Talk**   [*]Email: jalali@ucla.edu

*Abstract*—High throughput real-time instruments are needed to acquire large data sets for detection and classification of rare events. Enabled by the photonic time stretch digitizer, a new class of instruments with record throughputs have led to the discovery of optical rogue waves [1], detection of rare cancer cells [2], and the highest analog-to-digital conversion performance ever achieved [3]. Featuring continuous operation at 100 million frames per second and shutter speed of less than a nanosecond, the time stretch camera is ideally suited for screening of blood and other biological samples. It has enabled detection of breast cancer cells in blood with record, one-in-a-million, sensitivity [2]. Owing to their high real-time throughput, instruments produce a torrent of data - equivalent to several 4K movies per second - that overwhelm data acquisition, storage, and processing operations. This predicament calls for technologies that compress images in optical domain and in real-time. An example of this, based on warped stretch transformation and non-uniform Fourier domain sampling will be reported.

*Keywords—Optical data processing; Ultrafast imaging; Big data compression.*

## I. INTRODUCTION

Telecommunication systems routinely generate, capture and analyze data at rates exceeding billions of bits per second. Interestingly, the scale of the problem is similar to that of blood analysis. With approximately 1 billion cells per milliliter of blood, detection of a few abnormal cells in a blood sample translates into a "cell error rate" of $10^{-12}$, a value that is curiously similar to the bit error rate in telecommunication systems. This suggests that data multiplexing, capture, and processing techniques developed for data communication can be leveraged for biological cell classification.

Time-stretch dispersive Fourier transform is a method for real-time capture of ultra wideband signals. It allows acquisition of single shot optical spectra continuously and at tens to hundreds of million frames per second. It has led to the discovery of optical Rogue waves [1] and, when combined with electro-optic conversion, to record analog-to-digital conversion performance [3]. Combination of the telecommunication technique of wavelength division multiplexing (WDM) and the time-stretch technique [4], the time stretch camera known as STEAM [5-11] has demonstrated imaging of cells with record shutter speed and continuous throughput leading to detection of rare breast cancer cells in blood with one-in-a-million sensitivity [11-15]. A second data communication inspired technique called FIRE is a new approach to fluorescent imaging that is based on wireless communication techniques [16]. FIRE has achieved real-time pixel readout rates one order of magnitude faster than the current gold standard in high-speed fluorescence imaging [16].

Producing data rates as high as one tera bit per second, these real-time instruments pose a big data challenge that overwhelms even the most advanced computers [17]. Driven by the necessity of solving this problem, we have recently introduced and demonstrated a categorically new data compression technology [18, 19]. The so called Anamorphic (warped) Stretch Transform is a physics based data processing technique that aims to mitigate the big data problem in real-time instruments, in digital imaging, and beyond [17-21]. This compression method is an entirely different approach to achieving similar functionalities as compressive sensing [22, 23] and is more amenable to fast real-time operation because it does not require iterative numerical computations.

## II. TIME STRETCH IMAGING

The basic principle of time stretch imaging (STEAM) involves two steps both performed optically. In the first step, the spectrum of a broadband optical pulse is converted by a spatial disperser into a rainbow that illuminates the target. Therefore, the spatial information (image) of the object is encoded into the spectrum of the resultant reflected or transmitted rainbow pulse (Figure 1). A one-dimensional rainbow is used to acquire a line-scan. 2D image is obtained by scanning the one-dimensional rainbow in the second dimension or by a two-dimensional rainbow. In the second step, the spectrum of the image-encoded pulse is mapped into a serial temporal signal that is stretched in time to slow it down such that it can be digitized in real-time [4]. This optically-amplified time-stretched serial stream is detected by a single-pixel photodetector and the image is reconstructed in the digital domain. Subsequent pulses capture repetitive frames. The laser pulse repetition rate corresponds to the frame rate and the temporal width of the pulses corresponds to camera's shutter speed (exposure time). The key innovations in STEAM that enable high speed real-time imaging are photonic time stretch for digitizing fast images in real-time and the optical image amplification for compensating the low number of photons collected during the ultra-short shutter time [24].

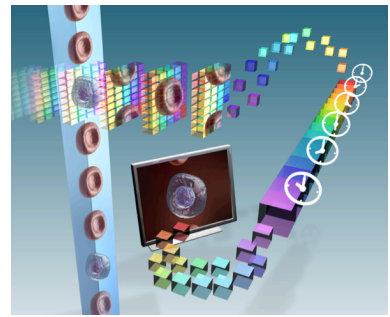

Fig. 1. Operating principle of time-stretch imaging. First, the object image is encoded in the spectrum of ultrafast optical pulses. Then, pulses are stretched in time by dispersive Fourier transformation, so that different wavelength components reach a single-pixel photodetector at different times. The time stretch function allows ultrafast image frames to be digitized in real-time. Images are optically amplified before detection and digitization to overcome the thermal noise.

### III. Cell Classification

Using time stretch imaging, we demonstrated high-throughput image-based screening of budding yeast and rare breast cancer cells in blood with an unprecedented throughput of 100,000 particles/s and a record false positive rate of one in a million [2]. Our first rare cancer cell detection method was based on imaging metal beads conjugated to cells expressing specific surface antigens [2]. However, when downstream operations such as DNA sequencing and subpopulation regrowth are desired, the negative impacts of biomarkers on cellular behavior are often unacceptable.

Subsequently, we demonstrated a label-free imaging flow-cytometry technique based on quantitative phase and intensity time stretch imaging of individual cells in flow [12]. The cell images are used to derive biophysical parameters such as size, total scattering, protein concentration, and uniformity. As an example, we have classified two different unlabeled cell types of mouse OT-II hybridoma T cells and human SW480 epithelial cancer cells with more than 90% accuracy. Figure 2 shows a three-dimensional scatter plot corresponding to diameter, protein concentration, and transparency used in classification.

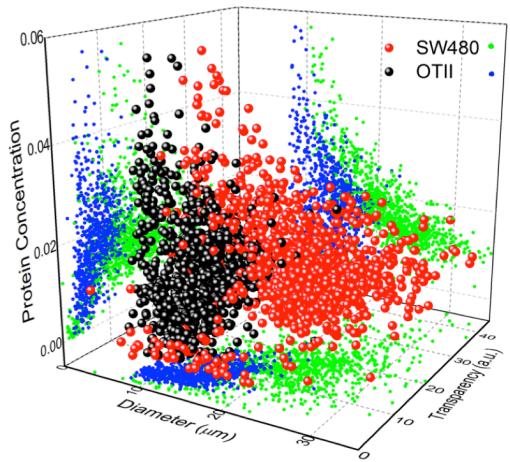

Fig. 2. Three-dimensional scatter plot based on size, protein concentration, and transparency of the cells measured by time stretch imaging, with two-dimensional projections for every combination of two parameters. The classification accuracies of more than 90% can be obtained from the hyper-dimensional cell information.

### IV. Nonuniform Fourier domain Sampling For Optical Data Compression

Using warped group delay dispersion, it has been shown that one can reshape the spectro-temporal profile of optical signals such that signal intensity's time-bandwidth product is compressed [17-21]. The compression is achieved through time-stretch dispersive Fourier transform in which the transformation is intentionally warped using an engineered group delay dispersion profile. This operation causes a frequency-dependent reshaping of the input waveform. Reconstruction (decoding) method depends on whether the information is in the spectral domain amplitude, or in the complex spectrum. In the time stretch camera, the image is encoded into the amplitude of the spectrum of a broadband optical pulse, and reconstruction consists of a simple nonuniform time-to-frequency mapping using the inverse of the warped group delay function.

To illustrate the concept in the context of time stretch imaging, we can consider a microscopic field of view consisting of a cell against a background such as a flow channel or a microscope slide (Figure 3). In the time stretch imaging, the object is illuminated by an optical pulse that is diffracted into a 1-D rainbow. This maps the 1-D space into the optical spectrum. The spectrum is then linearly mapped into time using a dispersive optical fiber with a linear group delay. The mapping process from space to frequency to time is shown in Figure 1a. The linearly stretched temporal waveform is then sampled by a digitizer resulting in uniform spatial sampling. This uniform sampling (also depicted in Figure 3a) generates redundant data by oversampling the sparse peripheral sections of the field of view. Such a situation evokes comparison to the mammalian eye where central vision requires high resolution while coarse resolution can be tolerated in the peripheral vision. In the eye, this problem is solved through nonuniform photoreceptor density in the retina. The Fovea section of the retina has a much higher density of photoreceptors than the rest of the retina and is responsible for the high resolution of central vision.

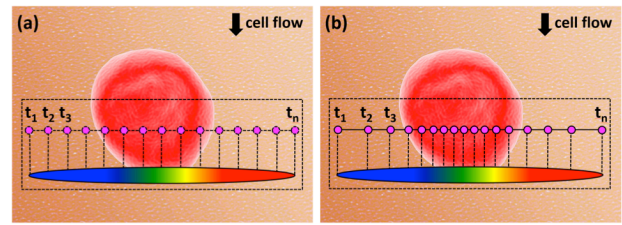

Fig. 3. Illustration of warped-stretch transform in imaging. (a) In conventional time stretch imaging (STEAM), the spectrum-to-time mapping is uniform, and the pixels are assigned equally-distanced across the field of view. (b) Using a nonlinear group delay profile in the spectrum-to-time mapping process results in a nonuniform sampling of the line image, assigning more pixels to the information-rich central part of field-of-view and less to the low-entropy peripherals.

We solve this problem by nonuniform mapping of spectrum into time via a warped group delay. The warped (anamorphic) space to frequency to time mapping is illustrated in the dotted box in Figure 3b. After uniform sampling in time, this leads to higher sampling density in the central region of the field of view and lower density in the sparse peripheral regions. The reconstruction is a simple unwarping using the inverse of the group delay.

V. RESULTS

The experimental setup used for our proof-of-principle demonstration of optical image compression consists of a STEAM system in which a fiber Bragg grating with customized chirp profile was used to achieve warped (anamorphic) stretch. The 2-dimensional image was reconstructed by stacking spectrally-encoded horizontal line images at different steps of the vertical scan (Figure 4a). The performance of this fiber Bragg grating was studied in [19] (Figure 4b).

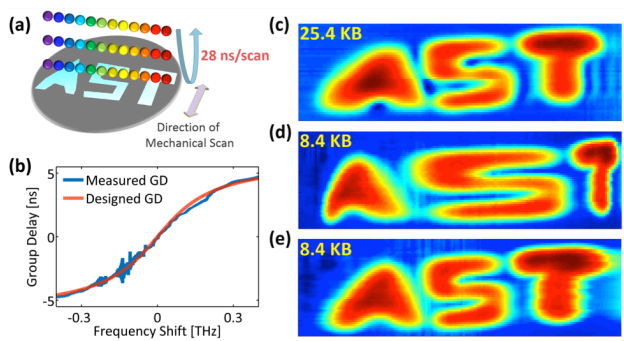

Fig. 4. Proof-of-concept experimental results. (a) The test sample reflected 1D rainbow illumination pulses, which are used to perform time stretch imaging at a scan rate of 36 MHz. (b) The warped-stretch transform leading to nonlinear spectrum-to-time mapping is performed by a custom chirped fiber Bragg grating with sublinear group delay (GD) profile. This profile gives higher group delay dispersion at the center frequency and reduced dispersion at the wings of the bandwidth. (c) If we use a linear group delay profile with the same dispersion as that of the warped-stretch at the center frequency, the reconstructed image size would be 25.4 KB. (d) The reconstructed image based on the waveform nonlinearly stretched by the chirped fiber Bragg grating has an obvious warping effect at the center of field of view (letter "S"). (e) The reconstructed image after unwarping is 8.4 KB achieving more than 3 times optical image compression.

If instead of the fiber Bragg grating, a dispersive fiber with linear group delay was used, the reconstructed image was as shown in Figure 4c. But, for the case of a fiber Bragg grating, since each line-scan is warped (anamorphic), the warping of the image is observed in the horizontal direction (Figure 4d). This effectively means that the central area (letter "S") is sampled with higher resolution than the peripherals (letters "A" and "T"). With the unwarping algorithm derived from the reverse dispersion profile, the uniform image was successfully reconstructed with a reduced data acquisition time and number of samples (Figure 4e). Compared to the case of the linear group delay (Figure 4c), an image with comparable quality is generated with only one-third of the data size (Figure 4e). We note that the reconstruction in this case is an intensity-only operation and does not require optical phase retrieval.

VI. CONCLUSION

Cross-over solutions from the world of telecommunication make it possible to create new types of real-time high-throughput instruments. These instruments have led to scientific discoveries, such as optical rogue waves and detection of rare cancer cells in blood. Storage, management, and processing of big data is the next challenge in real-time instruments. A new technique based on non-uniform Fourier domain sampling, made possible by warped (anamorphic) time stretch transformation, may solve this problem.